\documentclass[pre,onecolumn]{revtex4-2}

\usepackage{amsmath,amssymb,mathtools}
\usepackage{graphicx}
\usepackage{tikz}
\usepackage{pgfplots}
\usepackage{booktabs}
\usepackage{hyperref}

\newcommand{\up}{\uparrow}
\newcommand{\down}{\downarrow}

\begin{document}
	
	\title{Exact asymptotic equivalence between precisions of record-times and integrated currents}
	\author{Alberto Garilli}
	\email{alberto.garilli@unipd.it}
	\affiliation{Department of Chemical Sciences, University of Padova, via Marzolo 1, I-35131, Padova, Italy.}
	\date{\today}
	
	\begin{abstract}
		We establish an exact equivalence between the fluctuations of integrated currents and the temporal statistics of current record events in continuous-time Markov jump processes. Specifically, we show that, under minimal assumptions, the Fano factor of an integrated current coincides exactly with the squared coefficient of variation of the inter-record waiting times. This result is obtained by showing that inter-record times induce a renewal structure, since each record event resets the system to a fixed state, and that their generating function can be expressed exactly in terms of a combinatorial organization governed by Narayana numbers. The equivalence extends previous results obtained for unicyclic networks to arbitrary finite-state graphs and provides a unified representation of current fluctuations in terms of record-time observables. The results are derived within a purely probabilistic framework and clarify the structural origin of current precision in terms of first-passage-like event statistics.
		
	\end{abstract}
	
	\maketitle
	
	\section{Introduction}
	Fluctuations of currents are a defining feature of nonequilibrium stochastic systems \cite{seifert2012stochastic,seifert2018stochastic,derrida2007non}. They quantify precision, dissipation, and thermodynamic performance, and lie at the core of relations such as thermodynamic uncertainty relations \cite{seifert_barato_TUR,gingrich2016dissipation,pietzonka2017finite,horowitz2017proof}.
	Traditionally, these fluctuations are characterized through counting observables measured over long times \cite{touchette2009large,gingrich2016dissipation,wachtel2014fluctuating}.
	
	A complementary point of view emerges when one shifts the attention from counting to first-passage times.	An example is found in biochemical kinetics, where turnover times (the time between each production of some chemical species) provide direct access to noise properties of production currents, when the turnover reaction is irreversible \cite{schnitzer1995statistical,moffitt2014extracting} or when the reversed reaction exists but can not be observed \cite{cvexact}.	Moreover, it has been shown that recurrence and first-passage observables can encode thermodynamic information equivalently to current fluctuations \cite{gingrich2017fundamental,PhysRevResearch.3.L032034,10.21468/SciPostPhys.12.4.139,raghu2025thermodynamic}.
	
	A particularly striking observation, established for unicyclic Markov jump process, is that the fluctuations of an integrated current coincide with the precision of the times separating successive records of that same current \cite{wierenga2018quantifying}. More precisely, given the time-extensive non-equilibrium steady current $C_t$ with non-zero average, which counts the net number of crossings along a reference bidirectional channel, its Fano factor is
	\begin{equation}
		\mathcal{F}_C = \lim_{t\to\infty}\frac{\text{Var}[C_t]}{\langle C_t \rangle}.
		\label{eq:fano-factor}
	\end{equation}
	On the other hand, we may consider the inter-record times $\Delta T_m :=T_{m}-T_{m-1} > 0$, formally defined as the time between subsequent records of $C_t$, i.e. the difference between the first time $C_t$ assumes the value $m$ and the first time $C_t$ assumes the value $m-1$ in its historical record, assumed that $\langle C_t \rangle > 0$. The fluctuations of $\Delta T_m$ are quantified by its squared coefficient of variation
	 \begin{equation}
	 	\mathcal{P}_{\Delta T_m} = \frac{\text{Var}[\Delta T_m]}{\langle \Delta T_m \rangle^2}. 
	 	\label{eq:squared-precision}
	 \end{equation}
	 In Ref.\,\cite{wierenga2018quantifying} it was shown for unicyclic networks, where each record $M_t$ is directly associated to an increment in the winding number, that each instance $\Delta T_m$ induces a renewal structure on the sequence of records, resulting in them having identical distributions, i.e. $\Delta T_m \overset{d}{=} \Delta T$, and eventually leading to the exact identity
	 \begin{equation}
	 	\vert \mathcal{F}_C \vert = \mathcal{P}_{\Delta T}.
	 	\label{eq:equivalence-unicyclic}
	 \end{equation}
	 The absolute value extends the validity of the relation above to the case $\langle C_t\rangle<0$, provided that $M_t$ counts records in the direction of the typical growth of $C_t$: thus, for $\langle C_t\rangle>0$ ($\langle C_t\rangle<0$), $M_t$ counts only positive (negative) records of $C_t$, while records occurring in the opposite direction are disregarded.
	 
	 Eq.\,\eqref{eq:equivalence-unicyclic} extends the analogous relation valid when the transition contributing to the current $C_t$ is irreversible \cite{schnitzer1995statistical,moffitt2014extracting} and in the case where the transition is reversible but only one direction can be observed \cite{cvexact}. 
	
	In this work, we provide an explicit proof of the equivalence Eq.\,\eqref{eq:equivalence-unicyclic} extended to generic multicyclic networks with the only assumption of bidirectionality on the observed channel and hidden irreducibility \cite{singlecurrentFR}, i.e. irreducibility of the graph where the observed channel is removed. The result is obtained by adopting a transition space viewpoint \cite{PhysRevX.12.041026,van2022thermodynamic,singlecurrentFR,Garilli_2024}, where stochastic trajectories are punctuated by the individual forward and backward transitions contributing to $C_t$. In this way, $\Delta T$ can be decomposed as a combination of the times between these occurrences, known as inter-transition times \cite{PhysRevX.12.041026,van2022thermodynamic}. Interestingly, this reveals a precise combinatorial structure in Laplace space, connected with Narayana numbers \cite{petersen2015eulerian}, providing an exact expression of $\mathcal{P}_{\Delta T}$.
		
	An important consequence is that the asymptotic precision of a net current $C_t$ along a bidirectional channel can be evaluated in very general networks from finite-time measurements of the inter-record time provided that the observed dynamics is irreducible, and suggests a deep correspondence between a counting picture, based on integrated currents, and a temporal one, based on record events. A systematic comparison of the convergence properties of the statistical estimators associated with the counting and temporal pictures, aimed at determining whether either approach provides a statistical advantage, will be addressed in future work.

	\section{Setup}
	In this section we contextualize the result and set the notation to understand what follows.
		\begin{figure}
		\centering
		\includegraphics[width=\linewidth]{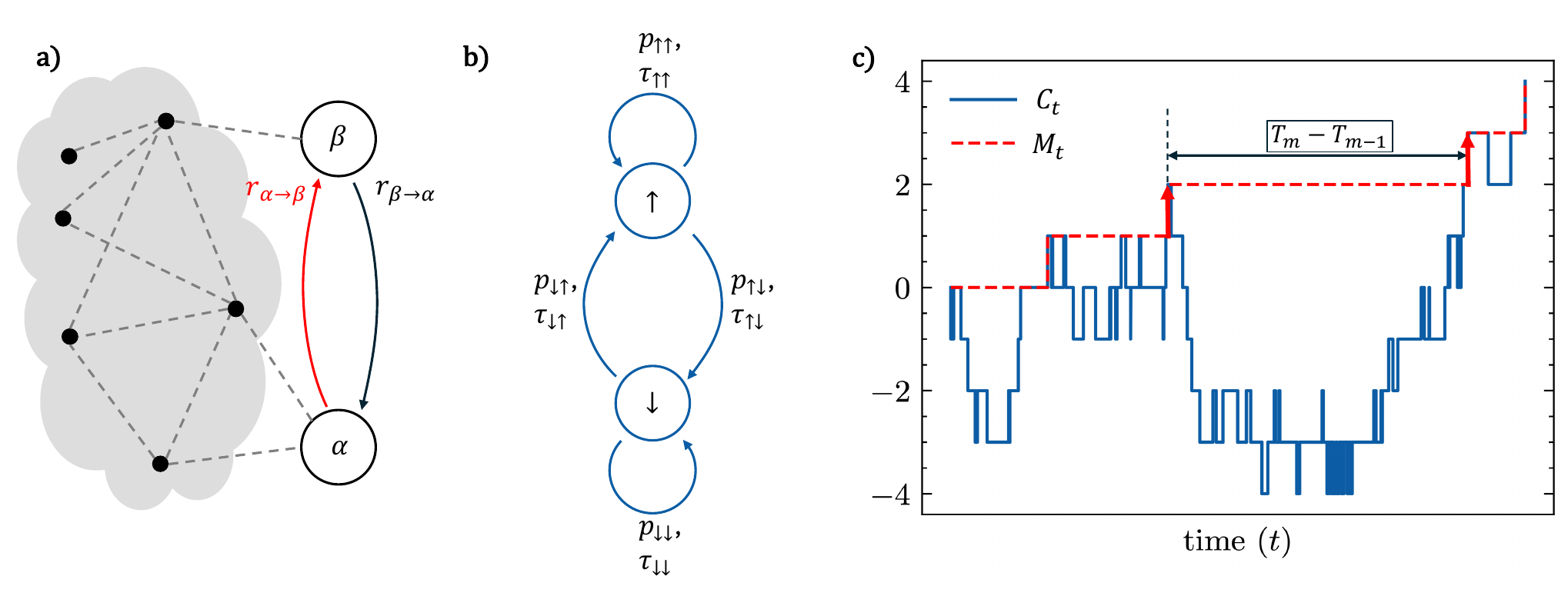}
		\caption{\textbf{a)} An example of a network of states connected by transitions. The bi-directional transition connecting states $\alpha$ and $\beta$ is observable, while we assume to have no knowledge about the hidden states and transitions (not necessarily bi-directional) occurring in the rest of the system (black dots and dashed lines inside the cloud). \textbf{b)} The associated trans-transition network. Each arrow indicates a possible succession of visible transitions, and it is associated with a trans-transition probability $p_{\ell'\ell}$ and an inter-transition time $\tau_{\ell'\ell}$. \textbf{c)} Example of a trajectory of the observable current $C_t$ along $\alpha\leftrightarrow\beta$. The evolution of the auxiliary inter-record process $M_t$ is highlighted by a red dashed line for the same realization of $C_t$. An inter-record time $T_{m}-T_{m-1}$ between the $m$-th and the $(m-1)$-th records of $C_t$ is highlighted.}
		\label{fig:network}
	\end{figure}
	
	\subsection{Continuous time Markov jump processes}
	We consider an irreducible continuous-time Markov jump process on a finite number of states such as the one in Fig.\,\ref{fig:network} a). A trajectory $\Gamma_t$ of lenght $t$ of the Markov process is defined as a succession
	\begin{equation}
		\Gamma_t = (x_0 , \tau_0 ) \longrightarrow ( x_1 , \tau_1 ) \longrightarrow \cdots  ( x_n , \tau_n )
	\end{equation}
	of states $x_i \in X$, $X$ denoting the state space of the process, with holding (or waiting) times $\tau_i$, i.e. the time spent in $x_i$ before jumping to $x_{i+1}$, which satisfy $\sum_i \tau_i = t$. Note that the intervals $\tau_0$ and $\tau_n$ are here understood as partial holding times, since the process may have entered $x_0$ before the beginning of the observation interval and may remain in $x_n$ after time $t$. Also note that the total number of jumps $n$ in the trajectory $\Gamma_t$ is itself a random variable.
	
	Let ${\rm p}_x(t) > 0$ denote the probability of finding the system at state $x \in X$ at time $t$ and collect all such probabilities in a vector $\boldsymbol{\rm p} (t)$. The time evolution is then governed by the Master Equation $\dot{\boldsymbol{\rm p}} (t) = -\boldsymbol{\rm R} \boldsymbol{\rm p} (t)$, with ${\rm R}_{xy} = -(1-\delta_{xy})r_{yx} + \delta_{xy}\sum_{z}r_{xz}$ indicating the rate matrix with jump rates $r_{yx} = r_{y\to x}$ from state $y$ to state $x$. Note that the sign convention on the rate matrix is opposite to the standard one, but is adopted here to simplify the subsequent analysis of inter-transition time statistics. From the irreducibility assumption, the Master Equation admits an unique stationary solution $\boldsymbol{\rm p}^{\rm ss}$ for each state $x$, with $\sum_x {\rm p}_x^{\rm ss} = 1$.
	
	We now focus on the pair of states $(\alpha,\beta)$ of $X$ and consider a particular bidirectional transition channel $\alpha \leftrightarrow \beta$ (possibly among several channels connecting the same pair of states). For a stochastic trajectory $\Gamma_t$, the anti-symmetric integrated current is defined as the net number of crossings along $\alpha\leftrightarrow\beta$ during the same time window	
	\begin{equation}
		C_t = \#_t(\alpha\to\beta)-\#_t(\beta\to\alpha),
		\label{eq:integrated-current}
	\end{equation}
	with $\#_t (\alpha\to\beta)$ and $\#_t(\beta\to\alpha)$ denoting the number of occurrences of $\alpha\to\beta$ and $\beta\to\alpha$ respectively during time $t$.
	 Assuming that the dynamics already reached stationarity at the initial observation time $t_0 = 0$, the average of the integrated current Eq.\,\eqref{eq:integrated-current} reads $\langle C_t \rangle = j t$, with $j = f_{\alpha\to\beta} - f_{\beta\to\alpha}$, where we identify the stationary fluxes $f_{x\to y} = r_{i\to y} {\rm p}_x$ from $x$ to $y$. Without loss of generality, we assume $j>0$, so that the transition $\alpha\to\beta$ is associated with positive increments of $C_t$. Crucially, the stationary fluxes satisfy $f_{x\to y} = 1/\langle \rho_{x\to y} \rangle$ \cite{frezzato2020stationary}, with $\rho_{x\to y}$ interpreted as the recurrence (or turnover) time, i.e. the time between successive occurrences of $x\to y$ regardless of how many times the inverse transition $y\to x$ occurred in between. 
		
	\begin{figure}
	\centering
	\includegraphics[width=0.6\linewidth]{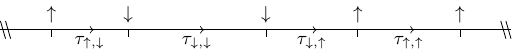}
	\caption{Graphical representation of the temporal succession $\up\to\down\to\down\to\up\to\up$ of visible transitions with the corresponding inter-transition times. The time-series of the total integrated current $C_t$ can be represented as a succession of visible transitions $\up$ (which increase $C_t$) and $\down$ (which decrease $C_t$). The time between two visible transitions $\ell'$ and $\ell$ is denoted by $\tau_{\ell',\ell}$.}
	\label{fig:timeline}
	\end{figure}
		
	\subsection{Reduced dynamics in transition space} 
	As stated above, the value of the current $C_t$ is only determined by the history of the forward and backward transitions along $\alpha \leftrightarrow \beta$. Rather than keeping track of the full-space trajectory $\Gamma_t$, we introduce a reduced description obtained by retaining only the occurrences of the transitions $\alpha\to \beta$ and $\beta\to\alpha$, together with the time elapsed between consecutive occurrences (see Fig.\,\ref{fig:timeline} for a graphical example). The resulting trajectory, denoted by $\tilde{\Gamma}_t$, lives in the space of visible transitions and fully specifies the evolution of the current $C_t$; its statistical properties can be naturally described within a transition-based framework. For brevity of notation, let $\ell$ denote a \emph{visible} transition, namely either $\up:=\alpha\to \beta$ or $\down:= \beta\to\alpha$, and let $\tau_{\ell\ell'}$ denote the inter-transition time \cite{PhysRevX.12.041026,van2022thermodynamic}, i.e. the time between the occurrences $\ell$ and $\ell'$ in a realization of the process, with no other visible transitions in between. $\tilde{\Gamma}_t$ is then the sequence
	\begin{equation}
		\tilde{\Gamma}_t = \overset{\tau_{\rm in}}{\longrightarrow} \ell_1 \overset{\tau_{\ell_1\ell_2}}{\longrightarrow} \ell_2 \overset{\tau_{\ell_2\ell_3}}{\longrightarrow} \cdots \overset{\tau_{\ell_{n-1}\ell_n}}{\longrightarrow} \ell_n \overset{\tau_{\rm fin}}{\longrightarrow},
	\end{equation}
	where the initial and final intervals $\tau_{\rm in}$ and $\tau_{\rm fin}$ do not belong to the sequence of inter-transition times. 
	
	In the remainder of this work, we additionally assume \textit{hidden irreducibility} of the underlying network for reasons that will be clarified later. This requirement consists in assuming the irreducibility of the graph obtained from the original one by removing the observed bidirectional channel between $\alpha$ and $\beta$ \cite{singlecurrentFR}. Under this condition, it is granted that any pair of consecutive events $\ell'\to\ell$, including $\up\to\up$ and $\down\to\down$, can occur with nonzero probability. Note that this condition does not require the hidden transition channels to be reversible; in fact, it only demands the existence of at least one hidden path from $\alpha$ to $\beta$ and one from $\beta$ to $\alpha$, without requiring that the two paths be reverses of each other. 

\begin{figure}
	\begin{tikzpicture}
		ff
	\end{tikzpicture}
\end{figure}

	\subsection{Trans-transition probabilities and moments of the inter-transition times}	
	\label{sec:moments}

	The moments of the inter-transition times are easily obtained from the detailed microscopic dynamics generated by the rate matrix $\boldsymbol{\rm R}$ \cite{PhysRevX.12.041026}. Precisely, they are expressed in terms of the \textit{survival matrix} $\boldsymbol{\textbf{K}}$ with $[\boldsymbol{\textbf{K}}]_{ij}= [\boldsymbol{\textbf{R}}]_{ij}$ for all $i,j$ except for $[\boldsymbol{\textbf{K}}]_{\alpha\beta} = [\boldsymbol{\textbf{R}}]_{\alpha\beta} + r_{\beta\alpha}^{(\nu)}$ and $[\boldsymbol{\textbf{K}}]_{\beta\alpha} = [\boldsymbol{\textbf{R}}]_{\beta\alpha} + r_{\alpha\beta}^{(\nu)}$, with $\alpha$ and $\beta$ connected by the observed bi-directional transition channel $\nu$, with rates $r_{\alpha\beta}^{(\nu)}$ and $r_{\beta\alpha}^{(\nu)}$. These expressions allow for multiple channels connecting the same pair of states, where only one channel is observed. Without loss of generality, we restrict to a single channel connecting $\alpha$ and $\beta$, leading to $[\boldsymbol{\textbf{K}}]_{\alpha\beta} = [\boldsymbol{\textbf{K}}]_{\beta\alpha} = 0$ This simplification is kept throughout the rest of this paper. In this way, the matrix $\boldsymbol{\textbf{K}}$ describes the dynamics of the system before it gets absorbed in either one of the observable transitions $\up:\alpha\to\beta$ or $\down:\beta\to\alpha$. Specifically, the joint probability density of an occurrence of $\ell \in \lbrace \up,\down\rbrace$ in the time interval $[t,t+dt)$, conditioned to a state $x_0$ is  \cite{singlecurrentFR}
	\begin{equation}
		p(t,\ell \mid x_0) = r_{\mathtt{s}(\ell)\mathtt{t}(\ell)} [e^{-t\boldsymbol{\textbf{K}}}]_{\mathtt{s}(\ell)x_0},
		\label{eq:density_unconditioned}
	\end{equation}
	where $\mathtt{s}(\ell)$ and $\mathtt{t}(\ell)$ denote the source and target states of $\ell$ respectively. By choosing $x_0$ as the target state of a visible transition $\ell' \in \lbrace \up,\down\rbrace$ we consider the Laplace transform of the density Eq.\,\eqref{eq:density_unconditioned}, which reads $\mathcal{L}_{\ell'\ell}(u) = r_{\mathtt{s}(\ell)\mathtt{t}(\ell)} (u \boldsymbol{\textbf{I}} + \boldsymbol{\textbf{K}})_{\mathtt{s}(\ell)\mathtt{t}(\ell')}^{-1}$, obtaining the four trans-transition probabilities as
	\begin{equation}
		p_{\ell'\ell} = \mathcal{L}_{\ell'\ell}(0) = r_{\mathtt{s}(\ell)\mathtt{t}(\ell)} [\boldsymbol{\textbf{K}}^{-1}]_{\mathtt{s}(\ell)\mathtt{t}(\ell')},
	\end{equation}
	each representing the probability that the next observed transition is $\ell$ given that the last observed one was $\ell'$, regardless of the time between the two occurrences (the same expression is also obtained via a simple time-integration of Eq.\,\eqref{eq:density_unconditioned}, since all the eigenvalues of $\boldsymbol{\rm K}$ are strictly positive \cite{PhysRevX.12.041026,singlecurrentFR}). The matrix element $[\boldsymbol{\rm K}^{-1}]_{xy}$ does in fact contain the contribution of all hidden paths leading from state $y$ to state $x$ \cite{singlecurrentFR}. Additionally, we can define a version of the density Eq.\,\eqref{eq:density_unconditioned} conditioned on the next occurring transition being $\ell$, obtaining the probability density
	\begin{equation}
		p(t \mid \ell',\ell) = \frac{p(t,\ell \mid \ell')}{p_{\ell'\ell}} =  \frac{[e^{-t\boldsymbol{\textbf{K}}}]_{\mathtt{s}(\ell)\mathtt{t(\ell')}}}{[\boldsymbol{\textbf{K}}^{-1}]_{\mathtt{s}(\ell)\mathtt{t(\ell')}}},
		\label{eq:density_conditioned}
	\end{equation}
	which characterizes the statistics of the inter-transition times $\tau_{\ell'\ell}:=\tau_{\ell'\to\ell}$, and satisfy that $\int_{0}^\infty p(t \mid \ell',\ell) dt = 1$ \cite{PhysRevX.12.041026}. From Eq.\,\eqref{eq:density_conditioned}, we thus obtain the first two moments of $\tau_{\ell'\ell}$, i.e. the times any two consecutive observable transitions $\ell'$ and $\ell$ as
	\begin{align}
		\langle \tau_{\ell'\ell} \rangle & = \frac{1}{p_{\ell'\ell} } \frac{d\mathcal{L}_{\ell'\ell}(-u)}{d u}\Big\vert_{u=0} =  \frac{[\boldsymbol{\textbf{K}}^{-2}]_{\mathtt{s}(\ell)\mathtt{t}(\ell')}}{[\boldsymbol{\textbf{K}}^{-1}]_{\mathtt{s}(\ell)\mathtt{t}(\ell')}}, \\
		\langle \tau_{\ell'\ell}^2 \rangle &= \frac{1}{p_{\ell'\ell} } \frac{d^2\mathcal{L}_{\ell'\ell}(-u)}{d u^2}\Big\vert_{u=0} =  \frac{2[\boldsymbol{\textbf{K}}^{-3}]_{\mathtt{s}(\ell)\mathtt{t}(\ell')}}{[\boldsymbol{\textbf{K}}^{-1}]_{\mathtt{s}(\ell)\mathtt{t}(\ell')}} .
	\end{align}
	
	\subsection{Record process}
	To conclude this section, we introduce the auxiliary process used to extend Eq.\,\eqref{eq:equivalence-unicyclic} to generic Markov jump processes on multicyclic networks. Let $M_t$ denote the strictly increasing process which follows the historical records of $C_t$ (see Fig.\,\ref{fig:network}) c), here called \textit{record process} for simplicity. More precisely, whenever $C_t$ attains the positive value $m$ (provided that $j>0$) for the first time along a trajectory $\Gamma_t$, $M_t$ is updated to $m$. At later times, $M_t$ remains constant until the next record $m+1$ is reached, regardless of $C_t$ reverting to lower values $c<m$ within the same time window. We then denote with $\Delta T_m$ the time interval between the $m$-th and the $(m-1)$-th increment of $M_t$ (\textit{inter-record time}). Assuming that the forward and backward transitions which contribute to $C_t$ connect an unique pair of states $\alpha$ and $\beta$, the system is found in the same state $\beta$ at each increment of $M_t$, so that record events induce a renewal structure, as proved in detail in the next section. This, together with the strong Markov property at the regeneration times $T_m$ implies that the time intervals $\Delta T_m$ are i.i.d, and therefore we can focus on the statistics of a single such interval. 
	
	\section{Result}
	
	We now enunciate and prove the result of this work, which establishes an exact equivalence between the fluctuations of a bidirectional integrated current along a chosen channel $\alpha\leftrightarrow\beta$ and the precision of the associated inter-record times in generic multicyclic Markov jump processes satisfying hidden irreducibility with respect to the removal of the chosen channel.
	
	Consider a stationary Markov jump process, and let $C_t$ denote the integrated current along $\alpha\leftrightarrow \beta$. Let $\Delta T_m$ be the inter-record times defined by the successive increments of the record process $M_t$ associated with $C_t$. Then, in the long-time limit, the fluctuations of the current are \textit{exactly} determined by the fluctuations of the inter-record times through the identity
	\begin{equation}
		\vert \mathcal{F}_C \vert = \mathcal{P}_{\Delta T},
		\label{eq:main-result}
	\end{equation}
	where $\mathcal{F}_C$ is the Fano factor of the current, defined by Eq.\,\eqref{eq:fano-factor}, and $\mathcal{P}_{\Delta T}$ is the squared coefficient of variation of the inter-record time distribution, defined by Eq.\,\eqref{eq:squared-precision}. The absolute value accounts for the sign convention of the current; without loss of generality we assume $j>0$, while the case $j<0$ follows straightforwardly.
	
	The result holds for arbitrary multicyclic networks under the sole assumptions that the observed channel is bidirectional and that the underlying dynamics satisfies hidden irreducibility, ensuring the accessibility of all sequences of visible transitions. Importantly, it does not rely on any unicyclic structure \cite{wierenga2018quantifying} nor on an effective reduction to a dominant cycle, and therefore applies to genuinely multicyclic systems.
	
	The derivation of Eq.\,\eqref{eq:main-result} proceeds by reformulating the stochastic dynamics of the current $C_t$ in terms of sequences of visible transitions $\ell \in \{\uparrow,\downarrow\}$, and relies on the fact that successive inter-transition times are independent random variables (although not necessarily identically distributed), in the sense that the duration of any interval does not depend on the durations of preceding ones. This property implies that both recurrence times and inter-record times can be decomposed into sums of elementary inter-transition times, allowing their Laplace transforms to factorize accordingly. Building on the expression of the current Fano factor in terms of recurrence-time precisions obtained in Ref.\,\cite{garilli2026using}, we express both quantities within the same inter-transition framework. The resulting representations can then be directly compared, ultimately yielding the exact identity Eq.\,\eqref{eq:main-result}.
	
	\subsection{Independence of inter-transition times}
	\label{sec:independence}
	
	Let $(\ell_1,\ldots,\ell_n)$ be a sequence of visible transitions and let $(\tau_1,\ldots,\tau_{n-1})$ denote the corresponding inter-transition times, with $\tau_i := \tau_{\ell_i\ell_{i+1}}$. Since the occurrence times of visible transitions are stopping times of the underlying Markov jump process, the strong Markov property applies at each visible event. In particular, after the occurrence of a transition $\ell_i$, the future evolution depends only on the state reached by that transition, namely the target state $\mathtt{t}(\ell_i)$, and is independent of the previous history of the process.
	
	As a consequence, once a visible transition $\ell_i$ has occurred, the joint statistics of the next visible transition $\ell_{i+1}$ and of the corresponding inter-transition time $\tau_i$ are completely determined by $\ell_i$, irrespective of the sequence of visible transitions and inter-transition times preceding it. Therefore,conditioned to the first visible transition $\ell_1$, the joint density of the subsequent transitions and the associated inter-transition times factorizes as
	\begin{equation}
		p(\tau_1,\ldots,\tau_{n-1},\ell_2,\ldots,\ell_n \mid \ell_1)
		=
		\prod_{i=1}^{n-1}
		p(\tau_i,\ell_{i+1}\mid \ell_i).
		\label{eq:factorization_joint}
	\end{equation}
	
	Since the density $p(\tau_i,\ell_{i+1}\mid \ell_i)$ coincides with the joint density introduced in Eq.\,\eqref{eq:density_unconditioned}, the Laplace transform associated with a visible sequence also factorizes into a product of elementary contributions. Explicitly,
	\begin{equation}
		\mathcal{L}_{\ell_1\cdots\ell_n}(u)
		=
		\prod_{i=1}^{n-1}
		\mathcal{L}_{\ell_i\ell_{i+1}}(u),
		\label{eq:factorization_laplace}
	\end{equation}
	where $\mathcal{L}_{\ell_i\ell_{i+1}}(u)$ denotes the Laplace transform of the density $p(t,\ell_{i+1}\mid \ell_i)$ defined in Sec.\,\ref{sec:moments}.
	
	Alternatively, Eq.\,\eqref{eq:factorization_joint} can be equivalently written as
	\begin{equation}
		p(\tau_1,\ldots,\tau_{n-1}\mid \ell_1,\dots,\ell_n)
		=
		\prod_{i=1}^{n-1}
		p(\tau_i \mid \ell_i, \ell_{i+1}),
		\label{eq:factorization_conditioned}
	\end{equation}
	which shows that inter-transition times are independent once the sequence of visible transitions is fixed.
	
	The factorization property Eq.\,\eqref{eq:factorization_laplace} is the key ingredient in what follows. Since both recurrence times and inter-record times can be decomposed into sums of inter-transition times associated with suitable sequences of visible transitions, their Laplace transforms can be expressed in terms of products of the elementary Laplace transforms $\mathcal{L}_{\ell\ell'}(u)$. This provides a common framework for describing both classes of observables and ultimately for establishing the identity Eq.\,\eqref{eq:main-result}.
		
	
	\def\bigshift{40}
	
	\begin{figure}
		\centering
		\includegraphics{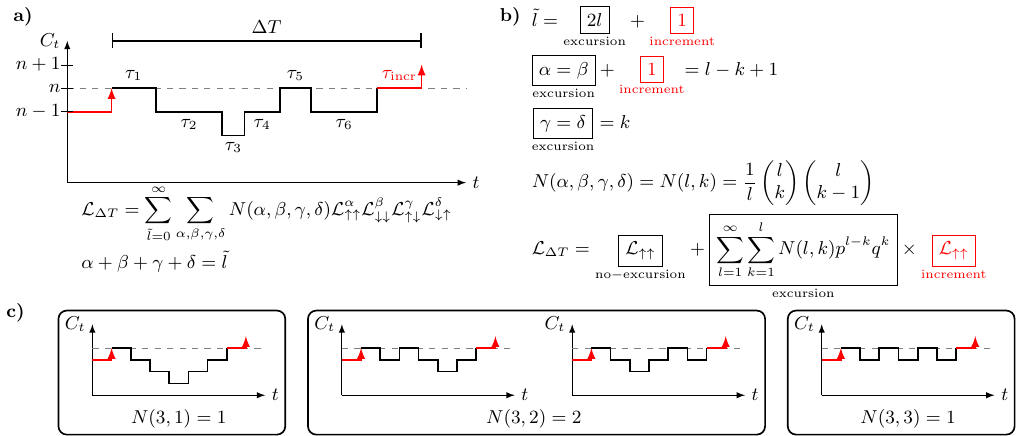}
		\caption{Construction of the Laplace transform Eq.\,\eqref{eq:laplaceDeltaT}. \textbf{a)} An excursion $\Delta T$ is defined as the time interval between two successive increments of the auxiliary process $M_t$ (red arrows). Each excursion is decomposed into elementary time intervals $\tau_i$, with $i=1,\dots,\tilde{l}$, where $\tilde{l}$ is the total number of intervals in the excursion. Each interval is identified with an inter-transition time, while the last one, $\tau_{\rm incr}$, is necessarily associated with the succession $\up\to\up$. Since the inter-transition times are independent, the Laplace transform of the distribution of $\Delta T$ factorizes into the product of the Laplace transforms of the individual inter-transition-time distributions. This yields the general expression shown, containing a combinatorial multiplicity $N(\alpha,\beta,\gamma,\delta)$, where, for brevity, $\alpha=\#_{\Delta T}(\up\to\up)$, $\beta=\#_{\Delta T}(\down\to\down)$, $\gamma=\#_{\Delta T}(\up\to\down)$, and $\delta=\#_{\Delta T}(\down\to\up)$. \textbf{b)} The relations between the exponents $\alpha$, $\beta$, $\gamma$, and $\delta$ are obtained by excluding the last time interval (the one eventually leading to the increment of $M_t$) and observing that \textit{i)} the remaining excursion contains an even number of intervals; \textit{ii)} the number of ``inversions'' (equivalently, peaks in the associated Dyck path \cite{petersen2015eulerian}) gives $\gamma=\delta=k$; and \textit{iii)} the numbers of consecutive increments and consecutive decrements of $C_t$ within the excursion are equal. These constraints imply that many distinct excursions carry the same weight in Laplace space, and that the number of excursions of length $2l$ with $k$ peaks is given by the Narayana numbers. \textbf{c)} All excursions with $l=3$ (i.e., excursions of length $6$) are shown, grouped according to the number of peaks $k$. Different durations of the elementary time intervals are not represented, as their contribution is already encoded in the Laplace transforms $\mathcal{L}_{\ell'\ell}$.}
		\label{fig:laplacetransf}
	\end{figure}
	
\subsection{Decomposition of the record process and Narayana numbers}
	
	The auxiliary process $M_t$ is updated whenever the integrated current $C_t$ attains a new historical maximum value $m$, and is therefore strictly increasing for $j>0$. With this convention, each increment of $M_t$ is associated with a visible transition of type $\up:\alpha\to\beta$, so that the system is always found in state $\beta$ immediately after a record event.
	
	The associated inter-record time $\Delta T_m = T_m - T_{m-1}$ can be decomposed into a sum of inter-transition times along a microscopic trajectory that starts and ends in the same state $\beta$. As discussed in Sec.\,\ref{sec:independence}, successive inter-transition times are independently distributed. Consequently, the inter-record times $\Delta T_m$ are independent; moreover, because each record increment is generated by portions of the full trajectory with the same initial and final state, they are also identically distributed. We may therefore focus on a single representative interval $\Delta T$.
	
	Using the same notation as in Sec.\,\ref{sec:independence}, the density of $\Delta T$ is a convolution of the probabilities Eq.\,\eqref{eq:density_unconditioned} under the constrain that $\sum_{i=1}^{n} \tau_i = \Delta T$, with $\tau_i$ the $i$-th inter-transition time between the two increments of $M_t$ and $n$ the total number of jumps in the interval $\Delta T$. Its Laplace transform is then factorized as a product of Laplace transforms as in Eq.\,\eqref{eq:factorization_laplace}. We now observe that between two consecutive increments of $M_t$, the excursions of $C_t$ always assume values smaller or equal to the current value $m$ of $M_t$. Let $\#_{\Delta T}( \ell'\to\ell)$ denote the number of times the sequence $\ell'\to\ell$ appears in the excursion of time duration $\Delta T$. Between two consecutive increments, the process may either perform a non-trivial excursion involving both visible transitions, or undergo a direct successive increment without any intervening $\down$ transition. In the case of a non-trivial excursion, the first transition in the excursion is necessarily $\down$, and the excursion terminates with two consecutive $\up$ transitions in order to increase $M_t$. In this case we have $\#_{\Delta T}(\up\to\down)=\#_{\Delta T}(\down\to\up)=k$, and $\#_{\Delta T}(\up\to\up)=\#_{\Delta T}(\down\to\down)+1=l-k+1$ (see Figure\,\ref{fig:laplacetransf} for a detailed explanation). If we exclude the last $\up$ transition, the problem is analogous to counting the number of excursions of lenght $2l$ with $k$ peaks (a peak is intended as a succession $\down\to\up$ inside the excursion). Notoriously, this is solved in terms of the Narayana numbers \cite{petersen2015eulerian}
		\begin{equation}
		N(l,k) = \frac{1}{l}  \left( \begin{matrix}
			l\\
			k
		\end{matrix}\right)  \left( \begin{matrix}
			l\\
			k-1
		\end{matrix}\right).
	\end{equation}	
	The Laplace transform of the inter-record time density of $\Delta T$ is then (see Fig.\,\ref{fig:laplacetransf} for an explanation)
	\begin{equation}
		\mathcal{L}_{\Delta T}(u) = \mathcal{L}_{\up\up}(u)\left(1 + S(u)\right),
		\label{eq:laplaceDeltaT}
	\end{equation}
	with
	\begin{equation}
		S(u)=\sum_{l=1}^{\infty}\sum_{k=1}^{l}
		N(l,k)\,p_u^{l-k} q_u^k,
		\label{eq:narayanagen}
	\end{equation}
	where $p_u \equiv p(u) = \mathcal{L}_{\up\up}(u) \mathcal{L}_{\down\down}(u)$ and $q_u \equiv q(u) = \mathcal{L}_{\up\down}(u)\mathcal{L}_{\down\up}(u)$. The series converges when $\sqrt{p_u} + \sqrt{q_u} \leq 1$. This condition is certainly satisfied for all $u\geq0$. In fact, in this regime the Laplace transforms decrease monotonically, implying $p_u < p_0$ and $q_u < q_0$, with $p_0 = p_{\up\up}p_{\down\down}$ and $q_0 = p_{\down\up}p_{\up\down}$ evaluated at $u=0$. Setting $a = p_{\up\up}$ and $b = p_{\down\down}$ for simplicity of notation
	\begin{equation}
		\sqrt{p_0} + \sqrt{q_0} = \sqrt{ab} + \sqrt{(1-a)(1-b)} = v\cdot w,
		\label{eq:cauchy}
	\end{equation}
	with vectors $v = (\sqrt{a},\sqrt{1-a})$ and $w = (\sqrt{b},\sqrt{1-b})$. By the Cauchy-Schwarz inequality, we conclude that $ \vert v \cdot w \vert \leq \Vert v \Vert \cdot \Vert w \Vert = 1$, and therefore $\sqrt{p_0} + \sqrt{q_0} \leq 1$, with equality holding for $a=b$, which corresponds to the case of vanishing effective affinity (see Eq.\,\eqref{eq:effective-affinity} below). 
	
	For $u<0$, it is sufficient to prove that the condition $\sqrt{p_u} + \sqrt{q_u} \leq 1$ is satisfied in a neighborhood of $u=0$, since we are interested in evaluating the derivatives of Eq.\,\eqref{eq:laplaceDeltaT} at $u=0$. Let $f_u = \sqrt{p_u} + \sqrt{q_u}$. When out of equilibrium (i.e. when $a\neq b$), the Cauchy-Schwarz inequality is strict, hence $f_0 < 1$. Since $f_u$ is continuous around $u=0$, there exist a neighborhood of the origin $u\in(-\epsilon,0)$ such that $f_u < 1$, which proves that the Narayana series remains convergent for $u<0$ sufficiently close to zero. The above argument cannot be applied when $a=b$, where $f_0 = 1$. However, this case corresponds to vanishing effective affinity and, consequently, to a vanishing average current. Since the Fano factor Eq.\,\eqref{eq:fano-factor} is only defined for non-zero currents, the (local) equilibrium case is excluded from the discussion.
	
	Ultimately, the series Eq.\,\eqref{eq:narayanagen} is the Narayana bivariate generating function with weights $p_u$ and $q_u$, and is explicitly expressed as \cite{petersen2015eulerian}
	\begin{equation}
		S(u) = \frac{1-p_u-q_u-\sqrt{(1-p_u-q_u)^2 - 4p_u q_u}}{2p_u}.
		\label{eq:narayana_function}
	\end{equation}
	Since $p_0 = p_{\up\up}p_{\down\down}$ and $q_0 = p_{\down\up}p_{\up\down}$,  with $p_{\ell'\up} + p_{\ell'\down} = 1$, $S_0 \equiv S(0)$ assumes the values
	\begin{equation}
		\begin{cases}
			S_0 =p_{\up\down}/p_{\up\up} \quad \text{when} \quad j>0,\\
			S_0 = p_{\down\up}/p_{\down\down} \quad \text{when} \quad  j<0,
		\end{cases}
		\label{eq:S0}
	\end{equation}
	so that $\mathcal{L}_{\Delta T}(0)$ is interpreted as the probability of ever increasing $M_t$ given that an increment just occurred. This problem can be formally rephrased in a more general way by also including a choice of the sign of $M_t$ opposite to the sign of the current's density $j$ as
	\begin{align}
		\mathcal{L}_{\Delta T} (0) &= \mathbb{P}[\exists \tau = \inf\lbrace t > 0 : C_t = 1 \rbrace < \infty \mid C_0 = 0, x_0 = \mathtt{s}(\up)],\nonumber \\
		& = \min \lbrace 1, e^A \rbrace ,
		\label{eq:simil-polettini}
	\end{align}
	where the minimum is chosen according to the sign of $j$ relative to $M_t$, and $A$ is the effective affinity
	\begin{equation}
		A = \ln\frac{p_{\up\up}}{p_{\down\down}},
		\label{eq:effective-affinity}
	\end{equation}
	which has the meaning of an effective force driving the local current along $\alpha\leftrightarrow\beta$ out of equilibrium. Note that Eq.\,\eqref{eq:simil-polettini} is analogous to a previously studied problem \cite{polettini2024multicyclic} when setting the initial probabilities to $p_1(\up) =  p_{\up\up}$ and $p_1(\down) = p_{\up\down}$, where $p_1(\ell)$ indicates the probability that the first observed transition is $\ell$ for a given initial observation time.	

	\subsection{Squared coefficient of variation of $\Delta T$}

	The squared coefficient of variation of the inter-record time $\Delta T$ is
	\begin{equation}
		\mathcal{P}[\Delta T] = \frac{\langle \Delta T^2 \rangle}{\langle \Delta T \rangle^2} - 1.
		\label{eq:cvDeltaT}
	\end{equation}
	Therefore, we compute the first two moments of the distribution of $\Delta T$ via its Laplace transform Eq.\,\eqref{eq:laplaceDeltaT} as
	\begin{equation}
		\langle \Delta T \rangle  = \frac{d \mathcal{L}_{\Delta T}(-u)}{d u}\Big \vert_{u=0}  \qquad \langle \Delta T^2 \rangle  = \frac{d^2 \mathcal{L}_{\Delta T}(-u)}{d u^2}\Big \vert_{u=0}.
	\end{equation}
	For brevity of notation, the dependence on $u$ is kept implicit, and we denote with $f'$ the derivative with respect to $u$ of $f(-u)$. Given the sign convention chosen for the rate matrix $\boldsymbol{\textbf{R}}$, we get, for the inter-transition time densities, that $\mathcal{L}_{\ell'\ell}' = (u\boldsymbol{\textbf{I}} + \boldsymbol{\textbf{K}})^{-2}$ and $\mathcal{L}_{\ell'\ell}'' = 2(u\boldsymbol{\textbf{I}} + \boldsymbol{\textbf{K}})^{-3}$, providing that, at $u=0$, $\mathcal{L}_{\ell'\ell}'(0) = p_{\ell'\ell} \langle \tau_{\ell'\ell}\rangle$ and $\mathcal{L}_{\ell'\ell}''(0) = p_{\ell'\ell} \langle \tau_{\ell'\ell}^2\rangle$ respectively. By differentiating Eq.\,\eqref{eq:laplaceDeltaT}, we obtain
	\begin{align}
	\mathcal{L}_{\Delta T}' & = \mathcal{L}_{\up\up}' \left( 1 + S \right) +  \mathcal{L}_{\up\up} S' \label{eq:first_moment} \\ 
	\mathcal{L}_{\Delta T}'' & = \mathcal{L}_{\up\up}'' \left( 1 + S \right)  + 2 \mathcal{L}_{\up\up}' S' + \mathcal{L}_{\up\up} S''. \label{eq:second_moment}
	\end{align}
	The derivatives of $S$ are easily found by noticing that the quantity in Eq.\,\eqref{eq:narayana_function} solves the quadratic equation
	\begin{equation}
		pS^2 + (p+q-1)S + q = 0,
	\end{equation}
	that provides, by differentiation, that
	\begin{align}
		S' &= - \frac{q' (S + 1) + p' S (S + 1)}{(2S + 1)p + q - 1} \label{eq:Sprime}\\
		S'' &= - \frac{2S'[q' + (2S+1)p' + S' p]+(S+1)(S p'' + q'')}{(2S + 1)p + q - 1}, \label{eq:Sprimeprime}
	\end{align}
	where, at $u=0$,
	\begin{align}
		p'_0 &= p_{\up\up}p_{\down\down}(\langle \tau_{\up\up}  \rangle  + \langle \tau_{\down\down}  \rangle ) \label{eq:p_prime} \\
		q'_0 &= p_{\up\down}p_{\down\up}(\langle \tau_{\up\down}  \rangle  + \langle \tau_{\down\up}  \rangle ) \label{eq:q_prime} \\
		p''_0 &= p_{\up\up}p_{\down\down}(\langle \tau_{\up\up}^2  \rangle  + \langle \tau_{\down\down}^2  \rangle + 2\langle \tau_{\up\up} \rangle\langle \tau_{\down\down} \rangle ) \label{eq:p_prime_prime} \\
		q''_0 &= p_{\up\down}p_{\down\up}(\langle \tau_{\up\down}^2  \rangle  + \langle \tau_{\down\up}^2  \rangle + 2\langle \tau_{\up\down} \rangle\langle \tau_{\down\up} \rangle ).  \label{eq:q_prime_prime} 
	\end{align}
	
	\subsubsection{First moment of $\Delta T$}
	Without loss of generality, we again restrict to $j>0$ and denote the forward and backward fluxes with $f_{\up} \equiv f_{\alpha\to\beta}$ and $f_{\down}  \equiv f_{\beta\to\alpha}$ to shorten the notation. From Eq.\,\eqref{eq:first_moment}, together with Eq.\,\eqref{eq:Sprime}, we evaluate the average of $\Delta T$ using the first expression in Eq.\,\eqref{eq:S0} for $S_0$ and Eqs.\,\eqref{eq:p_prime} and \eqref{eq:q_prime}. After a few passages we obtain
	\begin{equation}
		\langle \Delta T \rangle = \frac{1}{j},
		\label{eq:meanDeltaT}
	\end{equation}
	analogously to the unicyclic case \cite{wierenga2018quantifying}, and where we made use of \cite{garilli2026using}
	\begin{equation}
		\langle \tau_{\up\down} \rangle + \langle \tau_{\down\up} \rangle = \frac{p_{\up\down} + p_{\down\up}}{p_{\up\down}p_{\down\up}\omega} - \frac{p_{\up\up}}{p_{\up\down}} \langle \tau_{\up\up}\rangle - \frac{p_{\down\down}}{p_{\down\up}} \langle \tau_{\down\down}\rangle,
		\label{eq:relation_moments}
	\end{equation}
	with $\omega = f_\up + f_\down$ denoting the local dynamical activity, and the following relation between trans-transition probabilities and fluxes \cite{garilli2026using}
	\begin{equation}
	\frac{f_{\up}}{f_{\down}} = \frac{p_{\down\up}}{p_{\up\down}}.
	\label{eq:fluxes_ratio}
	\end{equation}
	
	Eq.\,\eqref{eq:meanDeltaT} is in accordance with the intuition that the rate of growth of $M_t$ is the same of $C_t$, quantified by the current's density $j = f_\up - f_\down$.
	
	\subsubsection{Second moment of $\Delta t$}
	
	For evaluating the second moment we combine Eq.\,\eqref{eq:second_moment} with Eqs.\,\eqref{eq:Sprime} and \eqref{eq:Sprimeprime}, using the first expression in Eq.\,\eqref{eq:S0} for $S_0$ and Eqs.\eqref{eq:p_prime} to \eqref{eq:q_prime_prime}. For the transition $\ell:\mathtt{s}(\ell)\to\mathtt{t}(\ell)$ we denote with $\bar{\ell}:\mathtt{t}(\ell)\to\mathtt{s}(\ell)$ its reverse transition. Hence, by additionally using that $p_{\ell'\ell} = 1-p_{\ell'\bar{\ell}}$, we obtain
	\begin{align}
		\langle \Delta T^2 \rangle = & a_1 \langle \tau_{\up\up}^2 \rangle + a_2 \langle \tau_{\down\down}^2\rangle + a_3 \left(\langle \tau_{\up\down}^2 \rangle + \langle \tau_{\down\up}^2\rangle \right) +  2 a_4 \langle \tau_{\up\up} \rangle \langle \tau_{\down\down} \rangle  \\
		& + 2a_5 \langle \tau_{\up\down} \rangle \langle\tau_{\down\up}\rangle + 2 a_6 \langle \tau_{\up\up}\rangle^2 + 2 a_7 \langle \tau_{\down\down}\rangle^2 + a_8 \left(\langle \tau_{\up\down} \rangle +\langle \tau_{\down\up} \rangle\right)^2 \\
		& + 2 a_9 \langle \tau_{\up\up} \rangle\left(\langle \tau_{\up\down} \rangle + \langle \tau_{\down\up} \rangle \right) + 2 a_{10}\langle \tau_{\down\down} \rangle \left(\langle \tau_{\up\down} \rangle + \langle \tau_{\down\up} \rangle \right),
		\label{eq:expanded1}
	\end{align}
	with coefficients $a_i$ depending on the trans-transition probabilities $p_{\ell'\ell}$, with $\ell,\ell' \in \lbrace \up,\down \rbrace$, and whose explicit expressions are reported in Appendix\,\ref{app:values}.

	\subsection{Decomposition of the recurrence times}
	
	In Ref.\,\cite{garilli2026using} an interesting relation between the Fano factor Eq.\,\eqref{eq:fano-factor} and the squared coefficients of variation $\mathcal{P}_{\rho_{\up}}$ and $\mathcal{P}_{\rho_{\down}}$ of the recurrence times $\rho_{\up}:=\rho_{\alpha\to\beta}$ and $\rho_{\down}:=\rho_{\beta\to\alpha}$ respectively was established. More explicitly, it was found that
	\begin{equation}
		\mathcal{F}_C = \coth \left(\frac{A}{2}\right) + \mathcal{P}_{\rho_\up} - \mathcal{P}_{\rho_\down},
		\label{eq:Fano_recurrences}
	\end{equation}
	with $A$ the effective affinity Eq.\,\eqref{eq:effective-affinity}. We exploit this relation together with the fact that recurrence times can also be expressed as combinations of inter-transition times to establish the main result of this work.
	
	We therefore consider the difference between the squared coefficients of variation of the forward and backward recurrence times appearing in Eq.\,\eqref{eq:Fano_recurrences}, which reads \cite{garilli2026using}
	\begin{equation}
	\mathcal{P}_{\rho_\up} - \mathcal{P}_{\rho_\down} = p_{\up\down} F_{\up}^2 (y + z(\up)) - p_{\down\up} F_{\down}^2 (y + z(\down)), \label{eq:diff_cvrecurrence}
	\end{equation}
	with
	\begin{equation}
		y = \langle \tau_{\up\down}^2 \rangle  + \langle \tau_{\down\up}^2 \rangle +  2\langle \tau_{\up\down} \rangle  \langle \tau_{\down\up} \rangle  + \frac{p_{\up\up}}{p_{\up\down}}\langle \tau_{\up\up}^2\rangle + \frac{p_{\down\down}}{p_{\down\up}}\langle \tau_{\down\down}^2\rangle ,
		\label{eq:y}
	\end{equation}
	independent of $\ell$, and
	\begin{equation}
		z(\ell) = 2\frac{p_{\up\up}}{p_{\up\down}} \langle \tau_{\bar{\ell}\bar{\ell}}\rangle (\langle \tau_{\up\down}\rangle + \langle \tau_{\down\up}\rangle) + 2\left(\frac{p_{\up\up}}{p_{\up\down}}\right)^2  \langle \tau_{\bar{\ell}\bar{\ell}}\rangle^2.
		\label{eq:z}
	\end{equation}
	
	\subsection{Proof of Eq.\,\eqref{eq:main-result}}
	
	To compare Eqs.\,\eqref{eq:cvDeltaT} and \eqref{eq:diff_cvrecurrence}, we use Eq.\,\eqref{eq:fluxes_ratio} to rewrite Eq.\,\eqref{eq:cvDeltaT} as
	\begin{equation}
		\mathcal{P}_{\Delta T} = \frac{F_\up^2}{p_{\down\up}^2} (p_{\up\up} - p_{\down\down}) \left(\langle \Delta T ^ 2 \rangle - \frac{1}{j^2}\right)
	\end{equation}
	and
	\begin{align}
		\mathcal{P}_{\rho_\up} -& \mathcal{P}_{\rho_\down} = \nonumber \\
		& F_{\up}^2\left( \frac{p_{\up\down}}{p_{\down\up}} (p_{\up\up}-p_{\down\down})y - p_{\down\up} z(\up) + \frac{p_{\down\up}^2}{p_{\up\down}}z(\down) \right),
	\end{align}
	with $y$ and $z(\ell)$ given by Eqs.\,\eqref{eq:y} and \eqref{eq:z} respectively. In this way, Eqs.\,\eqref{eq:cvDeltaT} and \eqref{eq:diff_cvrecurrence} become comparable: by taking the difference $\mathcal{P}_{\Delta T} - \mathcal{P}_{\rho_\up} + \mathcal{P}_{\rho_\down}$, we proceed by collecting similar terms. By applying Eqs.\,\eqref{eq:expanded1}, \eqref{eq:diff_cvrecurrence}, and considering that $p_{\ell'\ell} = 1-p_{\ell'\bar{\ell}}$, all terms with the second moments $\langle \tau_{\ell'\ell} \rangle$, for all $\ell,\ell' \in \lbrace \up,\down \rbrace$, vanish. The remaining terms provide
	\begin{align}
		& \mathcal{P}_{\Delta T} - \mathcal{P}_{\rho_\up} + \mathcal{P}_{\rho_\down} = \frac{f_{\up}^2}{p_{\down\up}^2} \left(\frac{p_{\up\up} + p_{\down\down}}{p_{\up\up} - p_{\down\down}}\right)  \times \\
		& \times \left[\gamma_1 (\langle \tau_{\up\down}\rangle + \langle\tau_{\down\up}\rangle)^2 + \gamma_2 \langle \tau_{\up\up} \rangle^2 + \gamma_3 \langle \tau_{\down\down} \rangle^2 + 2 \gamma_4 \langle \tau_{\up\up}\rangle \langle \tau_{\down\down}\rangle + 2\gamma_5 \langle \tau_{\up\up} \rangle (\langle\tau_{\up\down}\rangle + \langle \tau_{\down\up} \rangle) + 2\gamma_6 \langle \tau_{\down\down} \rangle (\langle\tau_{\up\down}\rangle + \langle \tau_{\down\up} \rangle)\right].
		\label{eq:expanded2}
	\end{align}
	The coefficients $\gamma_i$ depend on the trans-transition probabilities $p_{\ell'\ell}$ and are showed in Appendix\,\ref{app:values}. The expression above can be rewritten as
	\begin{align}
		\mathcal{P}_{\Delta T} - & \mathcal{P}_{\rho_\up} + \mathcal{P}_{\rho_\down} = \\
		& \frac{f_{\up}^2}{p_{\down\up}^2} \left(\frac{p_{\up\up} + p_{\down\down}}{p_{\up\up} - p_{\down\down}}\right) \left[p_{\up\down}(p_{\down\down}(\langle\tau_{\up\down}\rangle + \langle\tau_{\down\up}\rangle - \langle\tau_{\down\down}\rangle)-\langle\tau_{\up\down}\rangle-\langle\tau_{\down\up}\rangle) - p_{\up\up}p_{\down\up}\langle \tau_{\up\up}\rangle \right]^2,
	\end{align}
	which finally gives, by applications of Eqs.\,\eqref{eq:relation_moments} and \eqref{eq:fluxes_ratio}, that
	\begin{equation}
		\mathcal{P}_{\Delta T} - \mathcal{P}_{\rho_\up} + \mathcal{P}_{\rho_\down} = \frac{p_{\up\up} + p_{\down\down}}{p_{\up\up} - p_{\down\down}} = \coth\left(\frac{A}{2}\right),
		\label{eq:final-result}
	\end{equation}
	where we applied the definition Eq.\,\eqref{eq:effective-affinity} for the effective affinity $A$, and which, by comparison with Eq.\,\eqref{eq:Fano_recurrences}, finally implies the main result of this work
	\begin{equation}
		\mathcal{F}_C = \mathcal{P}_{\Delta T}.
	\end{equation}
	The proof can be repeated for the case $j<0$ by defining $M_t$ to count records in the direction of the typical growth of $C_t$, i.e., in the negative direction. The same steps lead to the general relation valid for all $j\neq 0$, which reads
	\begin{equation}
		\vert \mathcal{F}_C \vert = \mathcal{P}_{\Delta T}.
	\end{equation}

	\section{Discussion and conclusions}
	
	
	Beyond proving the equivalence itself, our analysis clarifies its structural origin. The key observation is that, although the hidden evolution between successive record events generally cannot be reconstructed from the observed trajectory and may involve arbitrarily complex excursions on the hidden network, each record returns the process to the same state, thus defining regeneration events: each record resets the system to the same state, so that successive inter-record times form a sequence of independent and identically distributed first-passage times. At the same time, expressing record times in terms of inter-transition times reveals a nontrivial combinatorial organization of trajectories in transition space, whose generating function is governed by Narayana numbers. From this perspective, the equivalence between current fluctuations and record-time fluctuations is no longer a remarkable coincidence, but rather the consequence of a common probabilistic structure underlying both observables.
	
	The present derivation also provides a unified framework encompassing all previously known instances of this correspondence. It naturally recovers the relation established for unicyclic networks \cite{wierenga2018quantifying}, where records coincide with increments of the winding number, as well as the case of irreversible observed transitions \cite{schnitzer1995statistical,moffitt2014extracting}. The latter includes both intrinsically irreversible transitions and reversible transitions for which only one direction is experimentally accessible \cite{cvexact}. In these situations, each record necessarily corresponds to the occurrence of a forward transition and the inter-record times reduce to the recurrence (or turnover) times considered in previous works. The present work shows that the same correspondence remains valid for arbitrary finite-state multicyclic networks with fully bidirectional observed transitions, the only additional assumption being hidden irreducibility, which guarantees that the effective affinity associated with the observed channel is well defined.
	
	Besides its conceptual significance, the equivalence also suggests a different perspective for characterizing current fluctuations. The precision of an integrated current is traditionally defined through asymptotic counting statistics \cite{touchette2009large,gingrich2016dissipation,wachtel2014fluctuating}, requiring the analysis of trajectories over increasingly long observation times. By contrast, the equivalent description derived here is formulated in terms of a sequence of independent inter-record times, reducing the problem to the statistics of independent first-passage observables which can be measured over finite times. This representation may be particularly advantageous in experimental situations where individual transition events are more naturally resolved than long-time current statistics. A notable example is provided by single-molecule experiments on molecular motors, where the experimentally observed quantities are the forward and backward mechanical steps \cite{svoboda1993direct,coppin1997load,nishiyama2002chemomechanical,carter2005mechanics,2023minflux,deguchi2023direct}, whereas the underlying chemomechanical dynamics is represented by a Markov jump process in which these steps correspond to transitions along a distinguished bidirectional channel $\alpha\leftrightarrow \beta$ \cite{PhysRevLett.98.258102,kolomeisky2007molecular}.
	
	More generally, the transition-space formulation adopted throughout this work provides a natural probabilistic language for describing nonequilibrium fluctuations in terms of observable events rather than complete microscopic trajectories. We expect that this viewpoint may prove useful in the study of more general classes of event-based observables and contribute to a broader understanding of the relation between counting statistics and first-passage descriptions in stochastic thermodynamics.
	
	\section{Acknowledgements}
	The author acknowledges D. Frezzato for carefully reading the first version of this manuscript and providing insightful comments. The financial contribution from “Fondazione Cassa di Risparmio di Padova e Rovigo” (CARIPARO) within the framework of the project ``NoneQ'', ID 68058 is also acknowledged.
	
	
	\bibliography{biblio}
	
	
	\appendix
	\section{Additional details}
	\label{app:values}
	
	\subsection{Details of Eq.\,\eqref{eq:expanded1}}
	Here we provide the explicit expressions of the coefficients $a_i$ appearing in Eq.\,\eqref{eq:expanded1} in the main text
	\begin{equation}
		\begin{aligned}
			a_1 & = \frac{p_{\up\up}p_{\down\up}}{p_{\up\up} - p_{\down\down}}, &
			a_2 & =  \frac{p_{\down\down}p_{\up\down}}{p_{\up\up} - p_{\down\down}} \\
			a_3 = a_5 & =  \frac{p_{\up\down}p_{\down\up}}{p_{\up\up} - p_{\down\down}}, &
			a_4 & =2  a_6 = 2 a_7 = 2\frac{p_{\up\up}^2 p_{\down\down} p_{\up\down} p_{\down\up}}{(p_{\up\up}-p_{\down\down})^3}\\
			a_8 & = \frac{2 p_{\up\up} p_{\up\down}^2 p_{\down\up}^2 }{(p_{\down\up}-p_{\down\down})^3}, &
			a_9  & = \frac{p_{\up\up}p_{\up\down}p_{\down\up}(p_{\up\up}-p_{\down\down}(1-2p_{\up\down}))}{(p_{\up\up} - p_{\down\down})^3}\\
			a_{10} = a_9 - a_3  &= \frac{p_{\down\down}p_{\up\down}p_{\down\up}(2p_{\up\up}p_{\up\down}+p_{\up\up}-p_{\down\down})}{(p_{\up\up} - p_{\down\down})^3}. &
			& 
		\end{aligned}	
	\end{equation}
	
	\subsection{Details of Eq.\,\eqref{eq:expanded2}}
	The explicit expressions of the coefficients $\gamma_i$ appearing in Eq.\,\eqref{eq:expanded2} in the main text are provided below
	\begin{equation}
		\begin{aligned}
			\gamma_1 & = p_{\up\down}^2 (1-p_{\down\down}^2) - 2_{p_{\down\down}}\frac{p_{\up\up}p_{\down\up}}{p_{\up\up} - p_{\down\down}}, &
			\gamma_2 & =  p_{\up\up}^2 p_{\down\up}^2 \\
			\gamma_3 & = p_{\up\down}^2p_{\down\down}^2 &
			\gamma_4 & = p_{\up\up}p_{\down\down}p_{\down\up}p_{\up\down} \\
			\gamma_5 & = p_{\up\up}p_{\up\down}p_{\down\up}^2 &
			\gamma_6  & = p_{\down\down}p_{\down\up}.
		\end{aligned}	
	\end{equation}
\end{document}